\documentstyle[10pt,aaspp4,flushrt]{article}
\def\etal{\it et al. \rm }
\begin{document}

\title{The Color-Magnitude Relation in Coma: Clues to the Age and Metallicity
of Cluster Populations}

\author{Andrew Odell}
\affil{Department of Physics and Astronomy, Northern Arizona University, Box 6010,
Flagstaff, AZ 86011; andy.odell@nau.edu}

\author{James Schombert}
\affil{Department of Physics, University of Oregon, Eugene, OR 97403;
js@abyss.uoregon.edu}

\author{Karl Rakos}
\affil{Institute for Astronomy, University of Vienna, A-1180, Wien, Austria;
karl.rakos@chello.at}

\begin{abstract}

We have observed three fields of the Coma cluster of galaxies with a narrow
band (modified Str\"omgren) filter system.  Observed galaxies include 31 in
the vicinity of NGC 4889, 48 near NGC 4874, and 60 near NGC 4839 complete to
$M_{5500}=-18$ in all three subclusters.  Spectrophotometric classification
finds all three subclusters of Coma to be dominated by red, E type
(ellipticals/S0's) galaxies with a mean blue fraction, $f_B$, of 0.10.  The blue
fraction increases to fainter luminosities, possible remnants of dwarf
starburst population or the effects of dynamical friction removing bright,
blue galaxies from the cluster population by mergers.  We find the color-magnitude
(CM) relation to be well defined and linear over the range of $M_{5500} =
-13$ to $-$22.  The observational error is lower than the true scatter around
the CM relation indicating that galaxies achieve their final positions in
the mass-metallicity plane by stochastic processes.  After calibration to
multi-metallicity models, bright ellipticals are found to have luminosity
weighted mean [Fe/H] values between $-$0.5 and $+$0.5, whereas low luminosity
ellipticals have [Fe/H] values ranging from $-$2 to solar.  The lack of CM
relation in our continuum color suggests that
a systematic age effect cancels the
metallicity effects in this bandpass.  This is confirmed with our age index
($\Delta(bz-yz)$) which finds a weak correlation between luminosity and mean
stellar age in ellipticals such that the stellar populations of
bright ellipticals
are 2 to 3 Gyrs younger than low luminosity ellipticals.  With respect to
environmental effects, there is a slight decreasing metallicity gradient with
respect to distance to each subcluster center, strongest around NGC 4874.
Since NGC 4874 is the dynamic and x-ray center of the Coma cluster, this
implies that environmental effects on low luminosity ellipticals are strongest
at the cluster core compared to outlying subgroups.

\end{abstract}


\section{INTRODUCTION}

The age and star formation history of elliptical
galaxies are, of course,
key tests to our understanding of galaxy evolution processes.  Our early
notions that ellipticals are old objects forming their stars at high
redshift in a single episode has come into
question by HST examination of
nearby dwarf galaxies (see review by Grebel 1998).  Detailed CMD's
indicate several episodes of star formation in dwarf galaxies, some
within a few Gyrs of the present.  The detection of Balmer lines (Caldwell
\etal 1993) and the high fraction of blue galaxies in distant clusters
(Rakos \& Schombert 1995) both challenge the monolithic collapse scenario
for ellipticals.

Observations of ellipticals from dwarf to giant in a range of environments
is the first step in resolving various predictions from hierarchical models
of galaxy formation.  In recent years, several spectroscopic programs have
determined the age and metallicity of Coma and Fornax galaxies through the
use of such indices as Mg$_2$ and H$\beta$ (Trager \etal 2000, Kuntschner
\etal 2001, Terlevich \& Forbes 2001, Poggianti \etal 2001).  Spectroscopic
studies are, of course, superior to determining metallicity values directly
from atomic lines.  While some comparison to models is required to interpret
the spectral age indicators, there are several independent estimators of both
age and metallicity.

An alternative approach is our narrow band photometry method, which focuses
on the shape of a galaxy's integrated spectral energy distribution (SED)
by way of near-UV and blue colors.
This has the advantage of measuring the behavior of the stellar population
as it reflects the temperature of the red giant branch and turnoff points,
but has the disadvantage of being very model dependent since the red
giant branch (RGB) must be
broad even in ellipticals and requires some estimation of the luminosity
weighted contribution from different metallicity populations (multi-metallicity
models).  In our most recent paper using the generalized Str\"omgren filter
system (Rakos \etal 2001), the possibility was explored of using various
color indices as age and metallicity estimators for ellipticals.  The integrated
colors of Galactic globular clusters, with known ages and metallicities, were
used to demonstrate and calibrate our narrow band system for single generation
objects.  Multi-metallicity models were derived from the globular
cluster (GC) calibration and
compared to the colors of dwarf ellipticals in Fornax.  While we have been
successful at predicting the global characteristics of ellipticals in clusters,
such as slope of the color-metallicity relation, with our simple multi-metallicity
models, our methods are by no means superior to spectroscopic survey but, in
fact, serve as a complement to their programs.  In general, spectral
investigations are superior in the detailed information they provide, but
the goal here is to develop a system that can be used for low brightness
objects or distant (i.e. faint) clusters where spectroscopy is very
impractical.

In an attempt to confirm and extend the Fornax results, we have carried
out photometry of three separate regions of the Abell cluster A1656 (Coma)
as a first step in the investigation of the behavior of low luminosity
ellipticals and dwarf galaxies in clusters of galaxies with increasing
redshift.  We have selected Coma because it is the nearest rich cluster
(richness class 2) with a dynamically evolved galaxy distribution
(Bautz-Morgan type II).  Coma is one of the most studied clusters in the
sky, and has long been regarded as the prototypical relaxed massive
cluster of galaxies.  However, recent studies of Coma have suggested that
the cluster is the product of a recent and an ongoing cluster-group merger.
While the bright galaxies in Coma are well understood, the distribution and
history of faint cluster galaxies has not yet been fully investigated.  In
addition, there has been a renewal of interest in the galaxy luminosity
function in rich clusters, in particular at its faint end. It is assumed that
low-luminosity galaxies and their evolution almost certainly play an important
cosmological role in explaining the large numbers of galaxies counted at
faint magnitudes.  Interest in the faint end of the luminosity function
derives from the fact that low luminosity ellipticals and dwarf galaxies
serve as key tests to our understanding of galaxy formation and evolution.
For example, in hierarchical models, galaxies are constructed from mergers
with smaller-mass dwarfs.

The goal of this paper is threefold.  First, we will examine the cluster
populations for the three regions in Coma around NGC 4889, 4874 and 4839
in terms of the spectrophotometric classifications.  Two recent papers
(Rakos, Odell \& Schombert 1997 and Rakos, Dominis \& Steindling 2001)
have shown that the ratio of the number of blue to red galaxies in
clusters (A2317 and A2218 respectively) has a strong dependence on
absolute magnitude, such that blue galaxies dominate at both the bright
and faint end of the cluster luminosity function.  We will examine if the
same effect occurs in Coma, a present-day counterpart to these
intermediate redshift clusters.  Second, we will investigate the behavior
of the color-magnitude (CM) relation for ellipticals.  In particular, we
will map the CM relation into our metallicity and age calibrations.
Lastly, we will examine the radial dependence of age and metallicity for
early-type galaxies around the three dominate galaxies to check for
environmental influences on the star formation history of cluster
galaxies.

\section{OBSERVATIONS}

\subsection{Data Acquisition and Reduction}

The Coma cluster (Abell 1656) is roughly 89 arcmins in radius, with three
dominant galaxies (NGC 4889, NGC 4874 and NGC 4839) forming the core of
three separate cluster components.  The first two of these, only 7 arcmins
apart, seem to be the relic central galaxies of previous groups that have
recently merged into the current cluster.  An additional subcluster,
centered on the third galaxy NGC 4839 40 arcmins to the southwest, seems
to be falling into the central region (Colles \& Dunn, 1996), or has just
passed through the main body of the cluster (Burns \etal, 1994), perhaps
triggering new star formation (Caldwell \& Rose, 1998).

Each section of the Coma cluster was observed with the Kuiper 1.55m
Telescope of the Steward Observatory on Mt. Bigelow Arizona.  Data was
taken with the LPL Focal Reducer and 2K CCD chip (binned to 1K) which
yielded 0.65 arc sec/pixel for a field size of 10.8 arcmin. We centered
one field on NGC 4889 (April 2-4, 2000) and the other on the central giant
NGC 4874 (May 10-11, 2000).  Our results are restricted to these central
regions and therefore are not necessary typical for the whole cluster. The
total April exposure time for the $uz$ filter was 11,700s, the $vz$ filter
9,600s, the $bz$ filter 3,000s and the $yz$ filter 3,000s. Each set of
exposures was divided in frames of 600s to 1200s to reduce the influence
of cosmic rays and irregularities of pixels on the chip.  For the May
observing run, the total exposure times were 7,200s in each of $yz$ and
$bz$ and 9600s in $vz$.  Weather conditions restricted the $uz$ filter to
2,400 seconds, and therefore it is of lower quality.  Calibration was
obtained through a number of spectrophotometric standards measured on each
night.  The region around NGC 4839 was observed with the 4-m Mayall
telescope of the National Optical Astronomy Observatory on Kitt Peak
Arizona (June 27-28, 2001). The Prime Focus T2KB CCD camera produces a 14
arcmin field with 0.42 arc sec/pixel.  The total exposure of 1,800s in
each filter consists of three 300s exposures each night.

The reduction procedures have been published in Rakos, Maindl \& Schombert
(1996) and references therein. The photometric system is based on the
theoretical transmission curves of filters (which can be obtained from the
authors) and the spectra of spectrophotometric standard stars published in
the literature (Massey \& Gronwall, 1990; Hamuy \etal 1994).  The
convolution of the transmission curves and the spectra of the standard
stars produces theoretical flux values for color indices of the standard
stars corrected for all light losses in the equipment and the specific
sensitivity of the CCD camera.  For distant clusters, the filters are
shifted for the proper redshift of the galaxies.  For Coma, the filters
were at zero redshift and a small k-correction was made.  The comparison
with the calculated theoretical fluxes delivers the corrections, which are
finally applied to the observed colors of galaxies with the same real
filters.  The disadvantage of the method is the need for a separate set of
real filters for different redshifts of clusters.  The main advantages of
this observing technique are that the filters are designed to sample
spectral regions which are very sensitive to changes in the underlying
physical properties and that these filters avoid all strong emission
lines, which cause confusion in UBV photometry of active galaxies.

Magnitudes were measured on the co-added images using standard IRAF procedures
and are based for brighter objects on metric apertures set at 32 kpc for
cosmological parameters of $H_o=75$ km s$^{-1}$ Mpc$^{-1}$ and the Benchmark
cosmology ($\Omega_m=0.3$, $\Omega_\Lambda=0.7$). For fainter objects the
aperture has been adapted to deliver the best possible signal to noise ratio,
but always using the same aperture for all four filters.  Color indices are
formed from the magnitudes: $uz-vz$, $bz-yz$, $vz-yz$, and $mz$ [$=(vz-bz)-(bz-yz)$].
Foreground galactic reddening in the direction of Coma is negligible and was
not taken into account.  Typical observing errors were 0.03 mag in color at
the bright end of the sample and 0.08 mag at the faint end.  A total number
of 415 objects were detected in all four filters.  A majority are rejected
as foreground stars or background galaxies based on our photometric criteria.
Some faint objects are rejected for insufficient signal in all four passbands
for accurate colors.

\subsection{Cluster Membership}

Cluster population statistics must deal with the contamination by stars as
well as by foreground and background galaxies.  Since our filters are
chosen to correspond to the redshift of the cluster, we can use the colors
to discriminate cluster membership in a very efficient manner.  We
calculated synthetic Str\"omgren colors for a compilation of approximately
150 spectra of nearby galaxies covering all possible morphological types.
It has been demonstrated that galaxies observed in their rest frame
Str\"omgren bands are confined to a very limited region of the
three-dimensional color space subtended by the four bands (Steindling,
Brosch \& Rakos 2001).  Since real galaxies were used as templates (and
not models), we need not worry about different constituents of model
galaxies.  The efficiency of this rejection mechanism is characterized by
the selected maximum deviation `sigma' between any measured color from the
same color of the best template galaxy.  Note that our filters are narrow
enough to obtain data of sufficient spectral resolution for identification
and classification purposes, but sufficiently wide so that the cluster's
velocity dispersion does not affect the colors.

Membership for the Coma cluster is straightforward for the bright end
since all galaxies down to $m_{5500}=16.5$ have measured redshifts.  This
comprises 53 galaxies of the 139 in the total sample.  Another 26 galaxies
fainter than 16.5 have measured redshifts from various deeper surveys.
For the remaining 60 galaxies, without redshifts, we assign membership based
on their colors and morphological appearance (i.e. most of these galaxies
are dE or dI systems, diffuse and low in surface brightness).  We note
that this sample is not complete in either magnitude or area.  We have
eliminated objects with poor photometry from the sample regardless of
apparent magnitude.  We are also focused on three fields surrounding NGC
4889, 4874 and 4839 with no attempt made to sample the outer region of the
cluster.  On the other hand, our sample has high reliability in the
sense that each member of the sample has a strong probability of being a
cluster member, and its spectrophotometric classification has a high
degree of confidence by only using high accuracy data (see next section).

\subsection{Spectrophotometric Classification}

Once cluster membership has been established, the principal moment values can
also be used to assign a spectrophotometric classification.  This
procedure is outlined in Steindling, Brosch \& Rakos (2001); however, that
project related photometric values with morphological appearance.  For our
goals, it is beneficial to relate the photometric values to a measure of
current star formation rate rather than morphological type.  Although we
find those photometric classifications will map into morphological ones
such that passive, red systems will typically be E/S0 types and
star-forming colors will typically be late-type spirals.

To this end, we have divided the first principal component axis (PC1) into
four subdivisions; E (passive, red objects), S (star formation rates
equivalent to a normal disk galaxy), S- (transition between E and S) and
S+ (starburst objects).  In addition, we can separate out objects with
signatures of non-thermal continuum (AGN) under the categories of A+, A
and A- to match their star forming colors.  It is important to remember
that these classifications are based solely on the principal components as
given by the color indices from four filters.  While, in general, these
spectrophotometric classes map into morphological ones (i.e. E types are
ellipticals, S- are S0 and early-type spirals, S are late-type spirals and
S+ are irregulars), the system has merit as an independent measurement
from morphology based on the color of the dominant stellar population in a
galaxy.  For Coma, some comparison to morphological types can be made.  In
all three subclusters, 72 of the galaxies have morphological types in the
literature.  Of these 72 morphological types, 63 are
spectrophotometrically classified as E type.  Within the 63 E types, 56
are classed as E/S0 morphologically. The remaining seven are classified as
S0/a or Sa.  Four S- type galaxies are S0's, two S galaxies are Sc's and 3
A types are E or S0's.  Unfortunately, Coma offers a very limited range
of morphological types; however, all the E types are early-type systems.
The seven Sa galaxies classified as E type are not of concern since we use
aperture magnitudes, which will heavily weight the large, red bulge in
early-type spirals.  Thus, regardless of the shape and appearance of a
galaxy, its spectrophotometric classification is based on the luminosity
weighted color of the underlying stellar population.

Following the membership procedure, Tables 1, 2 and 3 shows the final
selection of 139 cluster members for the NGC 4889 group, NGC 4874 group
and NGC 4839 group respectfully.  Final selection included all objects
with errors less than or equal to 0.08 mag.  Tables 1, 2 and 3 list the
position of the cluster objects (J2000), the observed colors and
magnitudes (apparent and absolute).  The derived quantities, described in
the text, are [Fe/H] and $\Delta(bz-yz)$ and the resulting galaxy class.
Note that [Fe/H] and $\Delta(bz-yz)$ are only calculated for E type
systems as restricted by our model interpretations.

\section{DISCUSSION}

\subsection{Characteristics of the Coma Cluster}

The Coma cluster is our nearest example of a rich, compact,
elliptical-rich system that is indicative of an evolved environment.  For
this reason, Coma also serves as a calibration for studies of rich distant
clusters and comparison of those cluster populations with a present-day
population.  However, unlike a classic cD cluster with a single dominant
elliptical at the bottom of the cluster potential surrounded by a
population of lesser cluster members, Coma offers a somewhat more diverse
cluster structure.  The cluster core is dominated by two equally bright
ellipticals NGC 4889 and NGC 4874.  Both have separate populations that
are visible in galaxy density diagrams (Geller \& Beers 1982) and x-ray
contours (Henry \etal 1981) centered on the ellipticals.  In addition,
Coma has a small subcluster to the southwest centered on NGC 4839.  These
three dynamically separate components contain a majority of the luminous
material in Coma.

The central ellipticals for the three subsystems of Coma also present a
diverse set of objects.  NGC 4889 and NGC 4874 have similar total
magnitudes (e.g. luminous mass), but differ in their structure.  NGC 4874
is a classic cD galaxies with an extended envelope typical of that class
of galaxies (Schombert 1987), whereas NGC 4889 is a D type elliptical
which are noted for their diffuse structure but lacking the extended
envelope of the cD class.  Dynamical studies indicate the origin of the
diffuse interior structure in D and cD galaxies is due to past mergers of
companion galaxies.  The origin of the extended cD envelopes is due to
cluster properties, as they relate to the ability for a cluster
environment to strip cluster members to develop a cluster-sized envelope
of stars that surrounds the central galaxy (Schombert 1988).

The Coma cluster provides a unique situation containing a cD galaxy, and
its extended envelope (NGC 4874) paired with a D galaxy and its
subpopulation (NGC 4889).  In addition, the southwest subgroup of Coma
centers on NGC 4839, itself an example of a low luminosity cD galaxy and
an example that the extended envelope phenomenon is associated with the
local cluster environment rather than the properties of the central
dominant galaxy.  Together the three subgroups within Coma provide a
laboratory to investigate the star formation history of galaxies in a
classic cD environment, a merger dominated D system and a small subcluster
with evidence of stripping to produce a low luminosity cD system.  We
begin this investigation an examination of the luminosity function for
each of these subclusters.

\subsection{Luminosity Functions}

The luminosity function of the Coma cluster has been well studied by
numerous groups over many wavelengths (see Smith \etal 1997 for
references).  Due to the narrow width of our filter system, our data is
not as deep in limiting magnitude as other surveys.  However, there are
sufficient numbers of galaxies to produce a luminosity function for the
populations around each giant elliptical.  Those values are shown for each
subcluster in the top panel of Figure 1 as log counts (N) versus
apparent magnitude ($m_{5500}$).  The unusual brightness of the first ranked
galaxy in each group is obvious in Figure 1.  This is a well known
phenomenon that supports the idea that first ranked ellipticals in rich
clusters owe about half their luminosity to the past mergers of lesser
cluster members (Schombert 1987).  The luminosity functions of the NGC 4889
and NGC 4874 subclusters have similar shapes, but the NGC 4874 group has
the denser concentration of galaxies, again indicating the fact that the
potential well is deeper around NGC 4874 compared to NGC 4889.  We note that
the faint end of the luminosity function for the NGC 4839 group rises more
sharply than the NGC 4889 or NGC 4874 groups.  This perhaps signals a more
advanced evolutionary state for the populations around NGC 4889 and NGC 4874
as dwarf galaxies are cannibalized to create their extended envelopes.

\begin{figure}
\plotfiddle{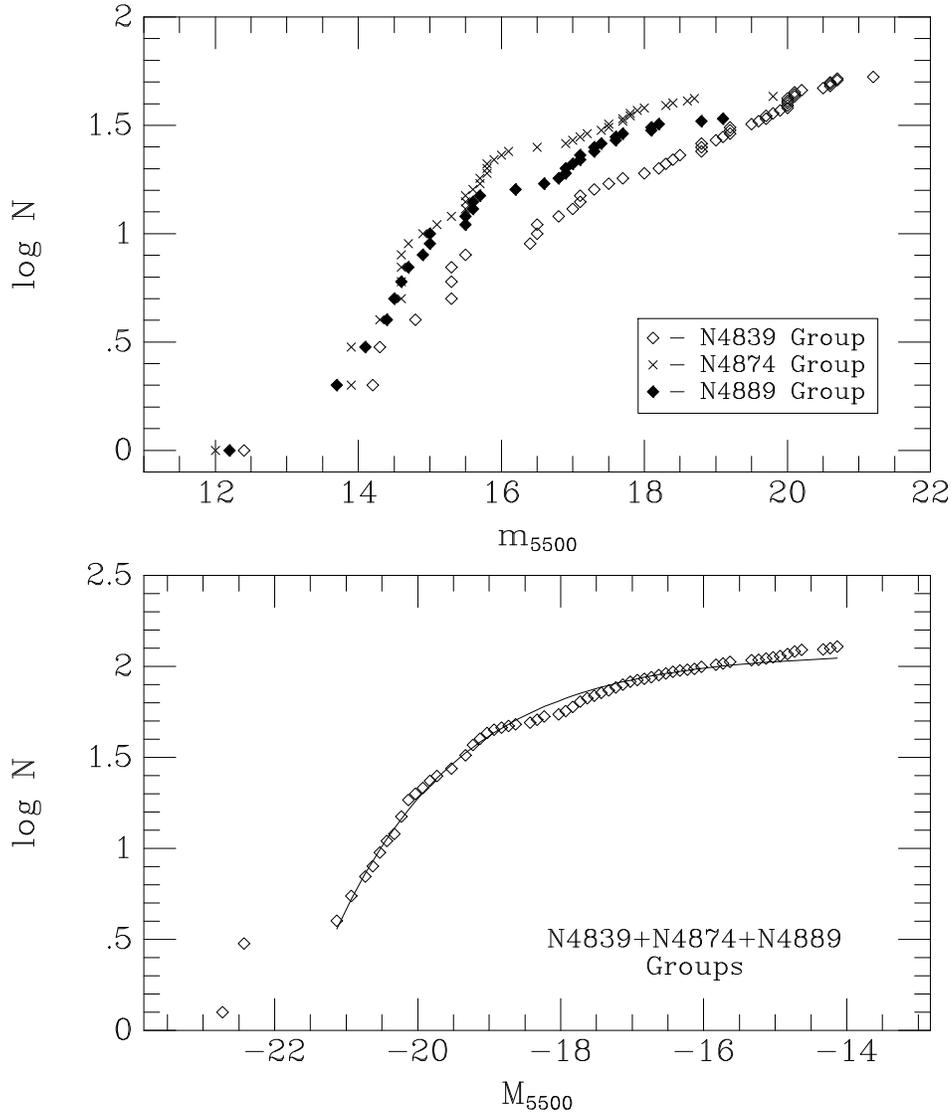}{11.0truecm}{0}{65}{65}{-200}{-30}
\caption{
The luminosity function for the three subclusters in
Coma.  The top panel displays the data for the individual subclusters, the
bottom panel is the summed population along with a best fit Schechter-type
function.  The total population shows the characteristic upturn at $m_{5500}=17$
($M_{5500} = -18$) with a the bright fit of $\alpha=1$ and the low luminosity
galaxies with a steeper $\alpha= 1.7$.  }
\end{figure}

The bottom panel in Figure 1 displays the summed luminosity function for
all three subclusters plus a single Schechter-type fit.  Despite the
limited number of galaxies we have measured, the luminosity function for
each population shows the characteristic upturn at $m_{5500}=17$
($M_{5500} = -18$).  Interestingly, most of the steepness at the faint end
is due to NGC 4839 group; however, the general trend of a break in the
luminosity function at $M=-19$ follows the results of previous studies
(Smith \etal 1997, De Propris \etal 1995, Secker \etal 1997, Thompson \&
Gregory 1993 and Bernstein \etal 1995).  The overall luminosity function
can be better fit by two Schechter-type components, the bright end having
the conventional $\alpha=1$, and the low luminosity galaxies with a
steeper $\alpha= 1.7$, similar to the characteristics of other rich
clusters such as Shapley 8 (Metcalfe \etal 1994).  Since field surveys
have demonstrated a shallow slope for the faint end ($\alpha=-1$, Ellis
\etal 1996), our narrow band luminosity function confirms the richer dwarf
population found in dense clusters.  Certainly, the population near the
bright galaxies in Coma is richer in low luminosity galaxies, as expected
from dynamical friction arguments (Schombert 1988).

\subsection{Butcher-Oemler Effect}

The signature test of galaxy evolution in cluster populations is the
Butcher-Oemler effect, the fraction of blue to red galaxies as a function of
redshift.  Being a photometric selection process, our classification scheme
makes for an obvious comparison to the Butcher-Oemler fraction.  In this
context, the population fractions, as defined by our spectrophotometric
criteria, are fairly typical of a low-redshift cluster.  For the total cluster,
we find that 71\% of the cluster members are E type (see Table 4), which
covers a majority of the galaxies that would be classified as E or S0
morphologically.  In comparison, for a rich cluster we expect the ratio
E:S0:Sp+Irr would be 20\%:40\%:40\% (Oemler 1992), slightly more spiral-rich
than our values.  However, we note our measurements are restricted to the
central part of the three densest subclusters and we would expect a larger
number of E/S0 galaxies in comparison to spirals due to the density-morphology
effect.  Table 4 presents the population fractions for the three subclusters.

The photometric change in cluster populations is measured by the fraction of
blue galaxies, $f_B$.  The original broadband definition of blue galaxies
$f_B$ given by Butcher \& Oemler (1984), is the ratio of galaxies 0.2 mag
bluer than the mean color of the E/S0 sequence (after k-corrections) to
the total number of galaxies in the cluster.  As discussed in Rakos,
Schombert \& Kreidl (1991), we have defined the fraction of blue galaxies
as the ratio of the number of galaxies bluer than $bz-yz=0.22$ to the
total number of galaxies, effectively the same prescription when
accounting for filter central wavelengths as the original Butcher \&
Oemler definition.  The relevant values of $f_B$ for each subcluster are
found in Table 4 as the number of blue galaxies.  The NGC 4889 group has
the highest number of blue galaxies (16\%) with the NGC 4874 group with
the least (2\%) and the NGC 4839 group in between (13\%).  The NGC 4839
data is slightly deeper in limiting magnitude, but selecting brighter
cutoff limits does not change the integrated $f_B$ values.  It is notable
that both NGC 4874 and NGC 4839 have cD envelopes providing circumstantial
evidence that dynamical environment effects are much stronger in those
subclusters than the NGC 4889 group.  The total $f_B$ value for the dense
regions of Coma is 0.10, a fairly standard value for nearby rich clusters
(Oemler 1992).

\begin{figure}
\plotfiddle{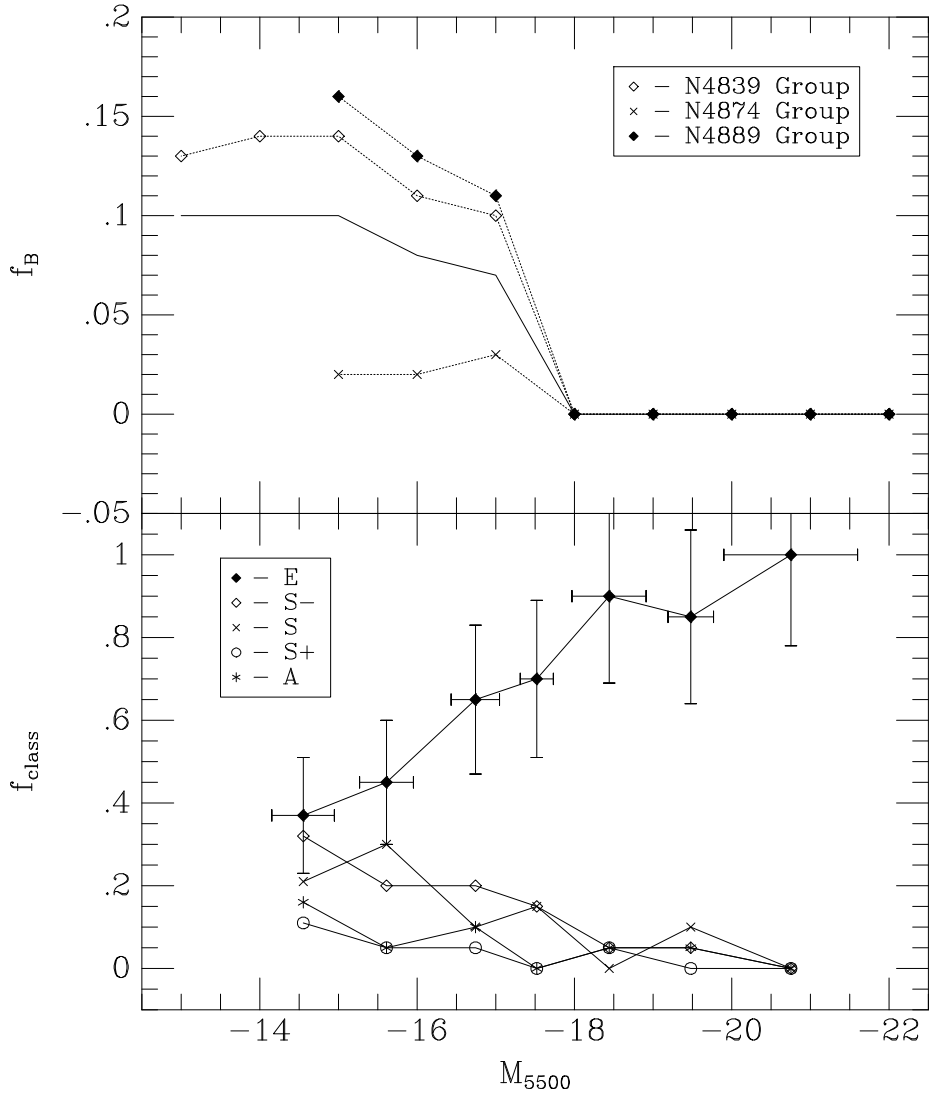}{13.0truecm}{0}{65}{65}{-200}{-30}
\caption{
The fraction of blue galaxies, $f_B$, as defined by the
Butcher-Oemler criteria, as a cumulative function of absolute luminosity.  The NGC
4874, NGC 4889 and NGC 4839 subgroups are marked as separate symbols, the
total cluster value is shown as the solid line.  The
trend of an increasing $f_B$ values with decreasing luminosity is obvious
for each subcluster in agreement with the result from distant clusters
(Rakos \etal 2001).  The low $f_B$ values for NGC 4874 and NGC 4839
correlate with the existence of a cD envelope for those systems.  NGC
4889, which lacks an extended envelope, has much higher $f_B$ values
indicating an environmental connection for the evolution of blue cluster
galaxies.  The bottom panel displays the spectrophotometric classifications in
bins of 20 galaxies per magnitude.  The blue galaxies are primarily due to
normal spiral-like star forming galaxies.  The decrease in E type as a
function of luminosity is primarily a metallicity effect.}
\end{figure}

The top panel in Figure 2 displays the cumulative behavior of $f_B$ with
luminosity for each of the three subclusters and the sum of the sample. As
is typical for rich clusters, the top of the luminosity function is
completely dominated by red, early-type galaxies (ellipticals and S0's).
As the luminosity drops below $-$18 we find a rise of $f_B$ due to the
increasing contribution from blue, low-mass galaxies.  Figure 2 is the
best example of how the interpretation of the blue fraction in clusters is
dependent on the completeness of observations with respect to cluster
boundaries and the photometric limiting magnitude since the the $f_B$
values do not stabilize until $M_{5500}=-15$.  This plateau is reached in
each of the various subclusters even though their final $f_B$ values vary
from 0.14 to 0.02.  In distant clusters, this issue becomes even more
salient as a contribution is found from a dwarf starburst population
(Rakos \etal 2000).  In fact, diagrams like Figure 2 are a key tool in the
interpretation of the evolution of blue galaxies since blue galaxies from
mergers of high mass, gas-rich systems will contribute to the top of the
luminosity function (Dressler \etal 1997); whereas a low mass starburst
population could influence the $f_B$ values for distant clusters than fade
and disappear in present-day clusters (Rakos \etal 2000).  In this context,
Coma displays none of the high luminosity blue galaxies found in
intermediate redshift clusters and the low luminosity blue galaxies have
colors similar to normal star-forming galaxies, rather than starburst or
AGN colors.

Lastly, the bottom panel of Figure 2 displays the run of spectrophotometric
classification as a function of absolute luminosity.  The distribution is
not cumulative, as in the top panel, and each bin contains 20 galaxies.  As
expected from the low $f_B$ values, a majority of the galaxies at high
luminosities are E type.  The few S types above $-$18 are red (early-type
spirals).  At fainter magnitudes, the fraction of E types declines and is
uniformly offset by the number of S- type of galaxies.  This is primarily a
metallicity effect, as the lower luminosity E type galaxies have bluer colors
due to low metallicities (see next section).

\subsection{Color-Magnitude Relation}

One of the first discovered relationships between the global properties
of galaxies and their underlying stellar populations is the
color-magnitude (CM) relation (see Bower, Lucey \& Ellis 1992 and reference
therein).  The relationship between redder galaxy color (both local and
global) with increasing luminosity is well known from the era of aperture
photometers (Sandage \& Visvanathan 1978).  The observed relationship in
luminous ellipticals has largely been interpreted as reflecting a cooler
RGB population (redder color) due to increased metallicity with larger
galaxy masses.  This has, in large part, been confirmed by numerous
absorption line studies (Trager \etal 2000) comparing line strengths of
features such as Mg$_2$ and Fe with dynamic measures of galaxy mass (i.e.
velocity dispersion) and luminosity (a corollary for constant $M/L$).
However, there is increasing evidence in the last few years that some of
the spread in the CM relation is due to age effects or recent star
formation, particularly at the low luminosity end (Poggianti \etal 2001).

Most CM relations found in the literature refer to the colors measured by
broadband filter systems versus total luminosity; however, Worthey (1994)
demonstrated that broadband colors suffer from age/metallicity
degeneracy.  Thus recent studies have focused on their spectral line
equivalents, line strength versus magnitude (Trager \etal 2000).  It is
usually considered an obvious point that spectral line measurements are
superior to colors, either narrow or broadband.  However, metallicity maps
into the colors of the underlying stellar population through the mean
temperature of the RGB, where metal-poor stars being hotter and,
therefore, bluer as compared to high metallicity stars.  Line strengths
directly measure these values (except where varying ratios, such as Mg/Fe,
complicate matters), but stellar populations vary according to the total
metallicity, $Z$.  Our narrow band colors form a useful compromise between
line studies and broadband colors through a measure of the continuum
shape.  In addition, spectral lines studies suffer from decreased S/N for
fainter objects, (i.e. high redshift or dwarfs) whereas our narrow band
system has been applied out to redshifts of 1.

\begin{figure}
\plotfiddle{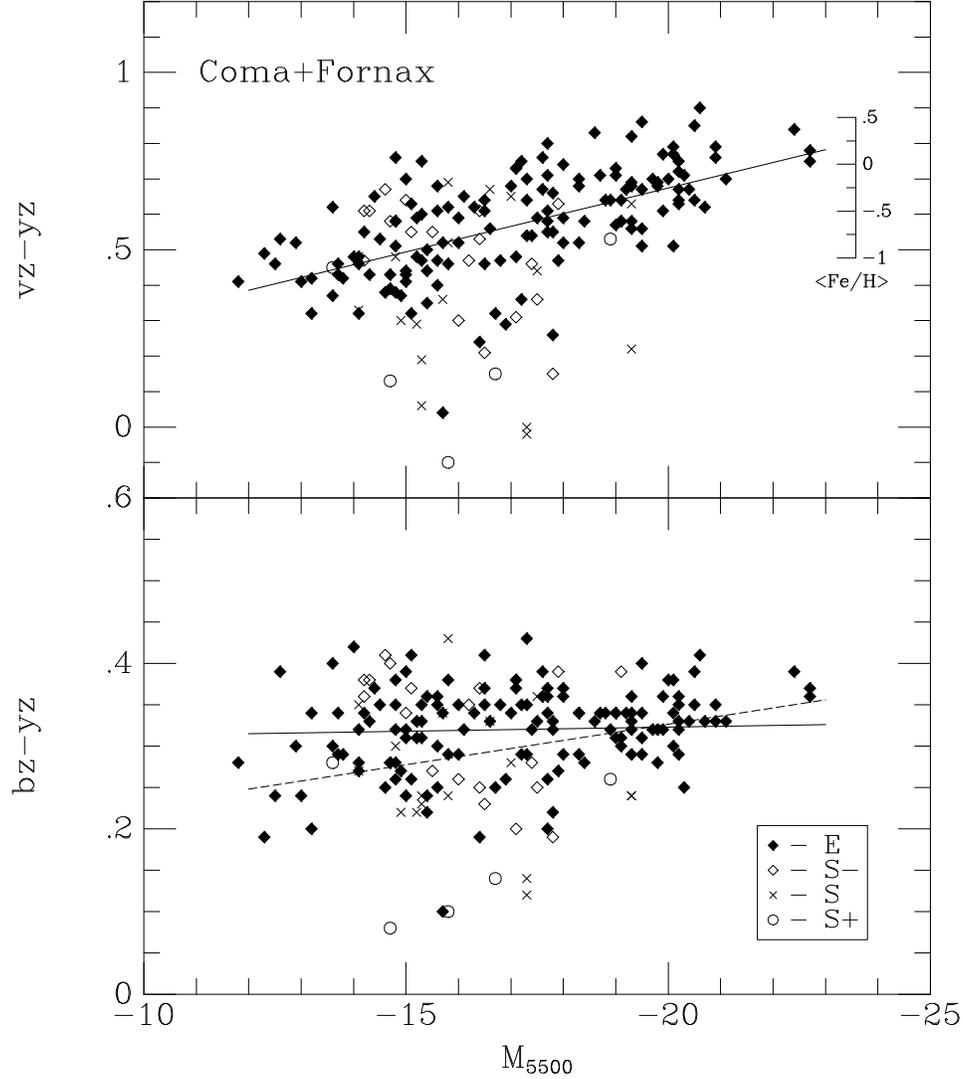}{13.0truecm}{0}{65}{65}{-200}{-30}
\caption{
The color-magnitude diagram for all three subclusters in the
Coma cluster.  The top panel display the metallicity color, $vz-yz$, as a
function of absolute magnitude, $M_{5500}$.  The bottom panel displays the
continuum color, $bz-yz$, as a function of apparent magnitude.  Also shown
is a solid line representing a robust least-squares fit to E type galaxies
only.  While the traditional color-magnitude relation is evident in the
$vz-yz$ colors, reflecting the mass-metallicity relation for galaxies,
there is no correlation visible in $bz-yz$.  The expected relationship,
base on multi-metallicity models, is shown as the dotted line.  The lack
of a relation indicates a systematic age effect.}
\end{figure}

The raw CM diagram, for all three Coma subclusters, is presented
in Figure 3 for the metallicity color $vz-yz$ (top panel), and the continuum
color $bz-yz$ (bottom panel).  The colors and magnitudes are based on
isophotal apertures, rather than weighted by surface brightness (unless the
galaxy is less than 16 kpc in radius in which case a total aperture magnitude
is used).  In general, Figure 3 contains all the major features for the
CM diagram for a diverse collection of galaxy types.  A main
ridgeline is found on the red side of the diagram that represents the sequence
of those galaxies classified both morphologically and photometrically as
ellipticals.  As noted in our photometry of distant clusters, the CM
effect is clearly visible for the $vz-yz$ color, yet the $bz-yz$ displays no
correlation with luminosity.  This is, in part, by design of the $uvby$ filter
system where, for composite populations, the $bz-yz$ is more sensitive to
contributions from hot stars (upper MS or blue HB populations) than from
changes in the RGB.  In contrast, the $vz-yz$ has a longer wavelength baseline
and several metallicity features in the $vz$ region to increase its sensitivity
to metallicity changes.

To determine the CM relation for old, non-starforming galaxies, Figure 3
displays as solid symbols E type galaxies (see \S 2.3) in Coma and a sample
of 47 dE galaxies from Fornax.  Fitting to an iterated least-squares regression
of color on magnitude produces the solid line in Figure 3 ($vz-yz = -0.036\pm0.005
M_{5500} - 0.046\pm0.084$).  This fit is indistinguishable in slope from
the color-magnitude relation determined from distant clusters (see Rakos
\etal 2000) out to redshifts of 0.3 with interesting variations in the
zeropoint to be discussed in a later paper.  Considering only the E types,
the scatter about the fit is greater than the observational error indicating
that, even among galaxies without current star formation, there is a range
of colors due to complicating factors such as reddening or differing mean
stellar age or simply that mass and metallicity do not map, linearly, into
the CM relation.  

Conversion of the color-magnitude diagram into a metallicity-mass relation
requires two steps, one model dependent, the other dynamical.  A dynamical
measurement is required to convert the luminosity of a galaxy into its
mass.  Lacking a companion or satellite for total dynamical mass
measurements, the standard procedure is to determine a mean $M/L$ from
rotation or velocity dispersion curves and apply that value to the total
luminosity of the galaxy.  We lack dynamical information for most of the
galaxies in our sample; however, ellipticals have an extremely narrow
range in $M/L$ and are well correlated with $L$ (Bernardi \etal 2002).
Thus, for our purposes, we have simply used the luminosity of the galaxies
($M_{5500}$) as a measure of mass with the assumption of constant $M/L$.

The conversion from $vz-yz$ color to metallicity (i.e. [Fe/H]) requires
the use of spectrophotometric models.  For a stellar population formed at
a unique time and metallicity (a single stellar population (SSP) such as a
globular cluster) the link between $vz-yz$ color and metallicity is
straightforward and documented in Rakos \etal (2001).  However, the
photometry used for our study is the total luminosity for each galaxy and
will be the sum of all the various stellar populations with a range of
metallicities and, perhaps, ages.  This is in contrast to spectroscopy
where the signal is weighted by surface brightness dominated by the 
stars in a galaxy's core and the assumption that a single stellar population
is observed will be close to valid.

In order to correct for the contribution of stars with a range of
metallicities, we have applied a very simple multi-metallicity approximation
of the underlying stellar population.  This method was described in Rakos
\etal (2001) and accurately reproduces the spectroscopic measurements of
metallicity with the expected colors.  In brief, the underlying metallicity
distribution is assumed to be Lorenzian in shape, a long tail to low
metallicities and a sharp cutoff at high metallicities.  A range of total
metallicities is constructed by holding the lowest metal population fixed
([Fe/H]=$-2.5$) and sliding the upper cutoff from high to low.  The Str\"omgren
colors for each metallicity bin are determined by empirical calibration to
globular clusters ([Fe/H]$<-1$) and
spectrophotometric SSP models ([Fe/H]$>-1$, Bruzual \& Charlot 2002).
The resulting metallicity scale is shown as an
inset in Figure 3.  With this calibration, and a $M/L$ ratio of 5 for early-type
galaxies, the CM relation maps into mass-metallicity as [Z/H] = $0.362\, {\rm
log}(M/M_{\sun}) - 3.736$.

In addition to plotting the CM relation for the metallicity color $vz-yz$,
data also exists to test its effect on the continuum color, $bz-yz$.  While
the continuum color has very few direct metallicity features to produce line
blanketening effects, the $bz-yz$ index is still sensitive to metallicity
changes primarily due to shifts in the color of MS turnoff stars.  Assuming
that a correlation should exist between absolute luminosity and $bz-yz$,
which is due solely to metallicity, then the Bruzual \& Charlot multi-metallicity
models, predicts a variation of 0.11 for the same range as the CM relationship
for $vz-yz$.  This model prediction is shown as a dotted line in the bottom
panel of Figure 3.  The data are not in agreement with this prediction despite
its excellent fit to the $vz-yz$ data, although there is a tendency for the
scatter to increase blueward for fainter magnitudes.  In fact, the $bz-yz$
index displays no correlation with galaxy luminosity.  This is not surprising
since our previous work with nearby ellipticals (Schombert \etal 1993) and
various distant clusters have all demonstrated no correlation between galaxy
luminosity and the $bz-yz$ index.

There are several effects that can redden $bz-yz$ while holding $vz-yz$ relatively
unchanged.  For example, reddening by dust can produce the effect seen herein;
however, it would require $A_V$ between one and two to cause the observed deviation
and this seems unlikely for normal cluster ellipticals based on their IRAS
luminosities.  Contributions from red and blue HB stars would reproduce
unusual MS turnoff colors, but again there is no expectation for such an
effect.  The most obvious candidate is an age effect, systematic with galaxy
luminosity.

\subsection{Mean Stellar Age}

Another advantage to narrow band continuum colors is that a crude index of
age can be assigned with the caveat that the galaxy being studied not have
any current, ongoing star formation.  As outlined in Rakos \etal (2001), a
measure of the mean age of a stellar population can be found by
comparing the
residual in the $bz-yz$ color with the expected value based on the mean
metallicity as determined by the $vz-yz$ color.  This is based on the fact
that the giant branch will be more sensitive to metallicity changes, whereas
the turnoff stars are more sensitive to age effects.

\begin{figure}
\plotfiddle{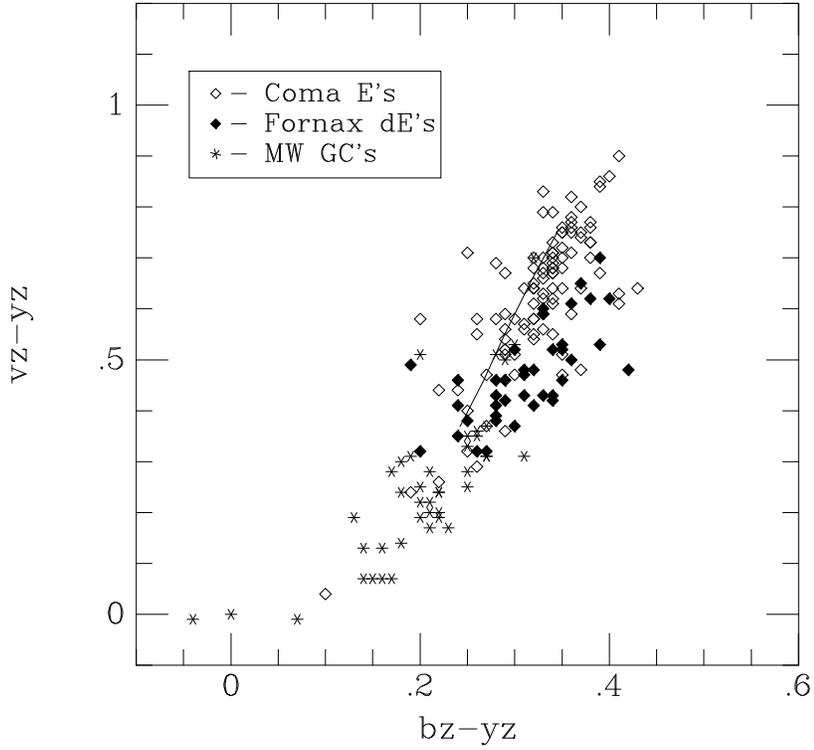}{8.0truecm}{0}{65}{65}{-200}{-110}
\caption{The two-color diagram for galaxies in the Coma cluster (open
symbols).  Also shown are dwarf ellipticals from the Fornax cluster (solid
symbols, Rakos, Dominis \& Steindling 2001) and globular clusters in our
own Galaxy (crossed symbols, Rakos, Dominis \& Steindling 2001).  Also
shown are the 13 Gyr models from Bruzual \& Charlot (2002).  A majority of
the low luminosity ellipticals in this diagram display $bz-yz$ colors that
are too red for their $vz-yz$, a sign of a redder turnoff population or
older mean stellar age.  }
\end{figure}

An advance indication of the results from the residuals in $bz-yz$ can been
seen in a two-color plot of the Coma E types, Fornax dwarf ellipticals and
globular clusters shown in Figure 4.  There appears to be a distinct population
of ellipticals that share a common region of the two-color diagram with dwarf
ellipticals from our Fornax work (Rakos \etal 2001).  Also shown in Figure
4 is the 13 Gyr model from Bruzual \& Charlot (2002) for a range of metallicity
(convolved to our multi-metallicity models).  It is clear from this comparison
that low luminosity and dwarf ellipticals are too red in $bz-yz$ for their
metallicity, indicating heavy reddening or an older mean stellar population.
This confirms the CM relation (or lack thereof) for $bz-yz$ as low luminosity
galaxies being too red for their $vz-yz$ colors.

The age indicator is the $\Delta(bz-yz)$ index, listed in Tables 1, 2 and 3
for the Coma E type galaxies.  In our Fornax study, $\Delta(bz-yz)$ was determine with
respect to a fiducial relationship as given by globular clusters with similar
ages.  Due to the much higher metallicities found for cluster ellipticals,
a comparison to metal-poor globular clusters is problematic as they occupy
such different regions of the two-color diagram.  As an alternative method,
we have calculated the residual $bz-yz$ color based on the distance in the
two-color diagram from the 13 Gyr models of Bruzual \& Charlot (2002).  These
models are an excellent fit to the red edge of the color-magnitude relation
(see Figure 3), and so provide a convenient initial guess at the age of elliptical
populations.  Since the $\Delta(bz-yz)$ index is a relative age scheme, the
choice of this fiducial zero point will not affect our results.

\begin{figure}
\plotfiddle{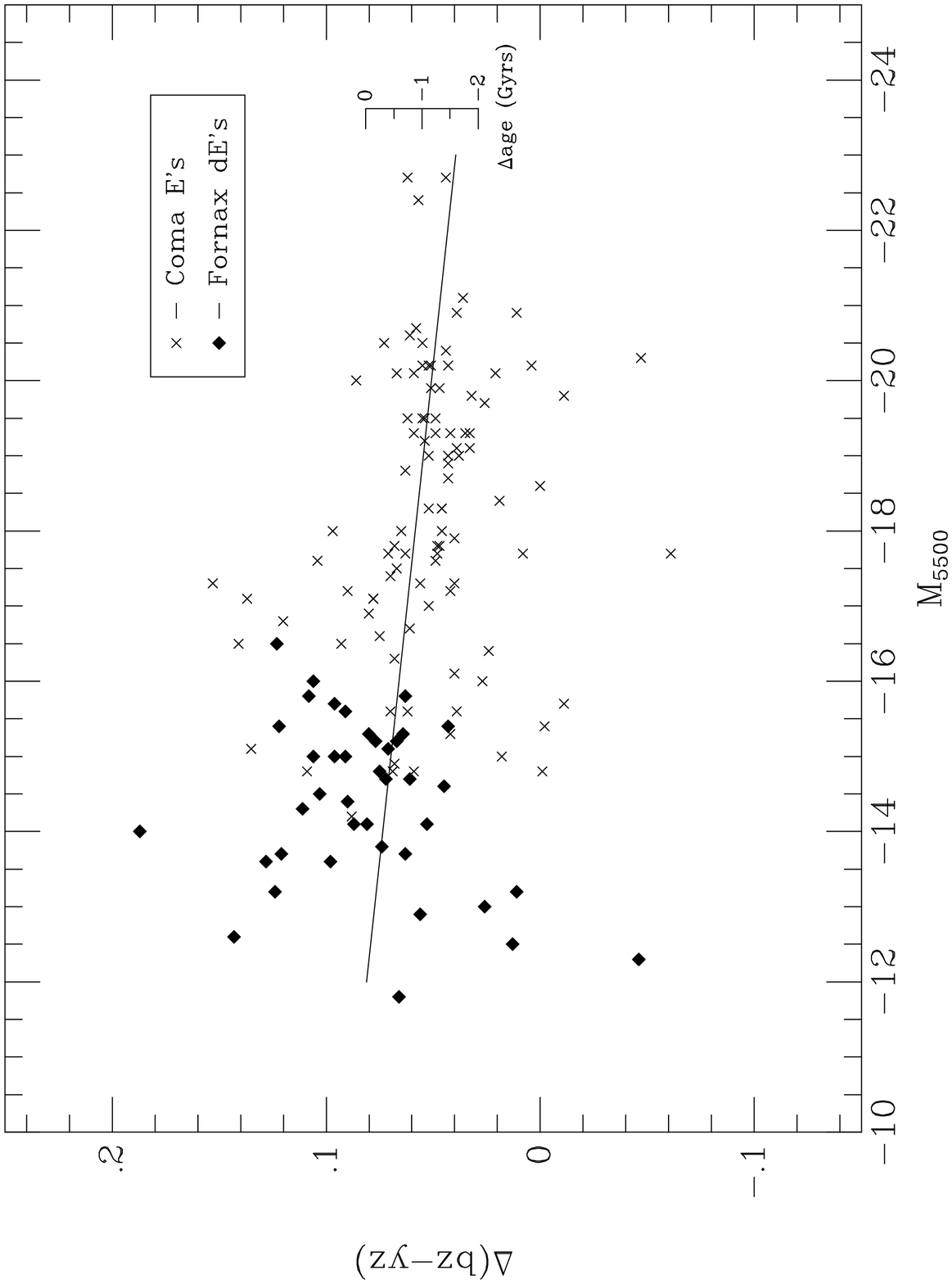}{9.0truecm}{-90}{65}{65}{-270}{360}
\caption{
The age indicator, $\Delta(bz-yz)$, versus absolute luminosity
for ellipticals in the Coma cluster (crossed symbols).  Also shown are the
data for dwarf ellipticals from the Fornax cluster (solid symbols, Rakos,
\etal 2001).  The dwarf ellipticals continue the trend of
redder $\Delta(bz-yz)$ values (older mean stellar age) seen in the Coma
data.  Also shown is a rough scale for the change in mean age of the stellar
population in Gyrs, relative to the dwarf elliptical sample. }
\end{figure}

Figure 5 plots the $\Delta(bz-yz)$ value as a function of luminosity for
the ellipticals in Coma.  Also included in the Figure, for comparison, are
the $\Delta(bz-yz)$ values for dwarf galaxies from Fornax (Rakos \etal
2001) recalculated to the Bruzual \& Charlot models.  The solid line is a
linear least squares fit to the Coma ellipticals which yields a slope of
0.0038$\pm$0.0018 with a correlation coefficient of 0.22.  The probability
of obtaining this or higher R by chance is 5\%.

As can be seen in
Figure 5, there is a clear trend for redder $\Delta(bz-yz)$ values with
lower luminosity for the Coma galaxies.  This implies the surprising
result that low luminosity ellipticals are older than their higher mass
counterparts.  As calibrated to globular clusters, a change in 0.25 in
$\Delta(bz-yz)$ corresponds to a change of 2 Gyrs in mean stellar age.
Thus, by a magnitude of $-$16, the mean age of a Coma elliptical is 2 Gyrs
older than the brighter ellipticals, assumed to be 13 Gyrs by their match
to the Bruzual \& Charlot 13 Gyr models.  This trend of older age with
fainter magnitude also merges nicely into the distribution defined by
Fornax dwarf ellipticals which have the reddest $\Delta(bz-yz)$
residuals.

Of course, our age indicator is a measure of mean stellar age for the
entire stellar population of each galaxy as weighted by luminosity.  Thus,
bright ellipticals can achieve a younger mean age through various
mechanisms.  The simplest explanation is, of course, that bright
ellipticals have later formation epochs than lower luminosity systems.
Various formation scenarios invoke the late formation of massive galaxies
(Kauffmann, White \& Guiderdoni 1993), but usually by assembling dwarf galaxies into larger systems.  If
dwarf galaxies have older stellar populations, then they are poor
candidates as the progenitors of massive ellipticals under this senario.
Another possibility
is that dwarf and massive ellipticals have the same formation epoch, but
massive ellipticals have a longer phase of initial star formation.  This
would produce a broad distribution of stellar ages, such that the mean age
is effectively younger than dwarf ellipticals whose initial star formation
timescale was brief.  This also has the attractive attribute that massive
ellipticals with a longer period of star formation also produce more
metals and a more enriched stellar population resulting in the observed CM
relation.  Lastly, since the ellipticals studied herein are members of a
dynamically evolved cluster, we cannot ignore the effects that mergers will
have on the cluster population, particularly the upper end of the
luminosity function.  One explanation to the decrease in the fraction of
blue galaxies for present-day clusters is that these galaxies are removed
from the cluster population by mergers.  The merger of galaxies with
recent star formation and younger stellar populations would lower the mean
stellar age as measured by our indices.  And this process would
preferentially occur more frequently for more massive ellipticals producing
the correlation seen in Figure 5.

\subsection{Environmental Effects on Metallicity and Age}

While our sample does not cover a large area within the Coma cluster, we
have sampled the three densest regions, those around the three brightest
cluster members.  This provides an opportunity to examine the run of
metallicity and age with distance from the central galaxy in each
subcluster.  Figure 6 plots the metallicity color, $vz-yz$, as a function
of radial distance from each of the brightest galaxies,
NGC 4839, 4874 and 4889.
There is a slight trend for decreasing metallicity with distance from the
core around NGC 4874. However, much of this correlation is blurred by the
large number of low luminosity galaxies nearby to the central galaxy,
presumably objects on orbits decaying by dynamical friction.  One concern
is that the weak trend with metallicity is due to a mass segregation
effect, where high mass galaxies with high metallicity occupy the
subcluster cores.  However, mass segregation is rarely seen in clusters
(Dominguez, Muriel \& Lambas 2001) and a check on absolute magnitude
versus distance displays no evidence of clustering of high luminosity
galaxies in the core (other than the fact that the D galaxies sit at the
cluster center).

\begin{figure}
\plotfiddle{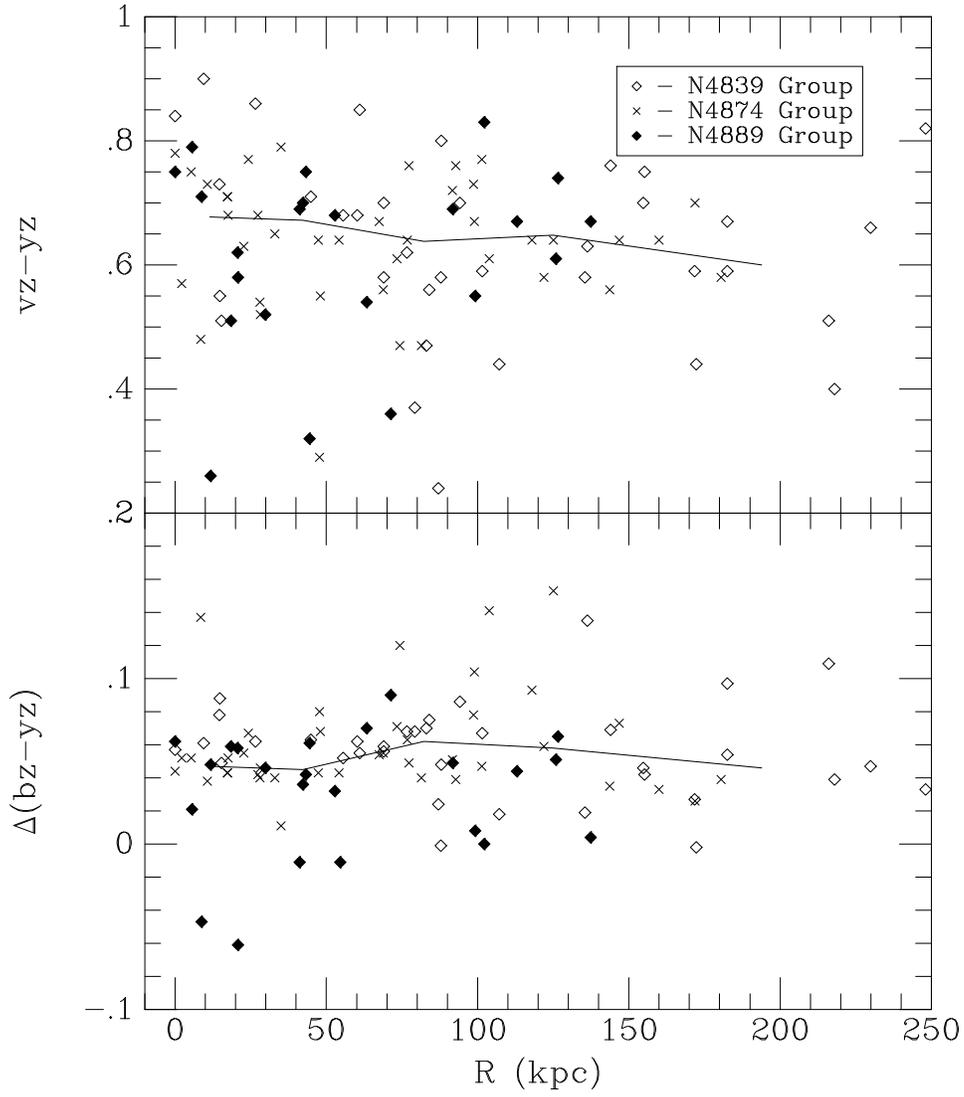}{13.0truecm}{0}{65}{65}{-200}{-30}
\caption{
Metallicity (top panel) and mean age (bottom panel) as a
function of cluster radius for the three Coma subclusters.  The primary
galaxies, NGC 4874, NGC 4889 and NGC 4839, form the positional center of
each group.  The solid lines display the running average.  While there is
a weak trend of decreasing metallicity with distance, there is no trend in
age for any group.}
\end{figure}

A trend of decreasing metallicity with radius is often found in the hot
x-ray gas in which clusters are embedded.  For example, Dupke \& White
(2000) find a factor of two change in the metallicity of the hot gas in Abell
496.  Since the x-ray gas in clusters derives from baryonic blowout of
ellipticals by galactic winds, the metallicity gradient in the gas is a
relic of the metallicity gradient in the galaxy distribution from early
epochs.  Comparison of the current galaxy gradients versus the x-ray
gradient should reveal the amount of dynamical and orbital evolution that
has occurred in rich clusters.

The trend between age and cluster radius is shown in the bottom panel of
Figure 6,
$\Delta(bz-yz)$ versus cluster radius.  There is no evidence of a radial
gradient for our age indicator.  This is somewhat surprising since both
metallicity and age correlate with luminosity, so we would expect age to
follow metallicity with respect to radial gradients.  However, the
correlation of age with luminosity is much weaker than metallicity and
luminosity, so the relic of a cluster age gradient may be lost in the
observational noise.

\section{CONCLUSIONS}

We can summarize our results in two parts, pure observational results and
interpretation.  From the observational side, we have found that the Coma
population is, unsurprisingly, rich in early-type galaxies based on their
spectrophotometric classification (71\%).  Morphologically, these systems
are predominately ellipticals and S0's as one would expect in a dense
cluster like Coma.  We summarize the analysis as the following:

\begin{itemize}

\item{} The fraction of blue galaxies rises slowly with decreasing
galaxy luminosity, reflecting the dynamically evolved nature of the Coma
system.  The blue fraction is lowest in those subclusters (NGC 4874 and
NGC 4839) with cD envelopes attached to the primary galaxy.  It seems
clear that the low blue fractions and the presence of a cD envelope are
due to stripping effects, on one hand removing stars for the development
of an expanded envelope while, on the other hand, halting star formation
by gas stripping and lowering the number of blue, star-forming objects.

\item{} What little blue population is found in Coma is unlike the
blue population in distant clusters.  It is neither bright, disk galaxies
nor faint starburst objects (i.e. there are no bright, blue galaxies in
Coma nor are there faint S+ type systems as found in A115 and A2283, Rakos
\etal 2000).

\item{} The color-magnitude relation for Coma (plus Fornax dE's) is
extremely well defined for E type systems from $M_{5500} = -23$ to $-$13,
although the scatter is larger than the observation error.  Our
metallicity color, $vz-yz$, maps into mean [Fe/H] such that the brightest
ellipticals have luminosity weighted mean [Fe/H] values of +0.3 while a
typical low luminosity elliptical ($M_{5500}=-15$) will have a [Fe/H]
value of $-$0.8.

\item{} While our metallicity color ($vz-yz$) displays a strong CM relation,
our continuum color ($bz-yz$) reveals no correlation with luminosity (i.e.
stellar mass), although there is a bluer increase in scatter to fainter
magnitudes.  This is unexpected since $bz-yz$ should be weakly sensitive
to metallicity through changes in the temperature of turnoff stars and
suggests a counteracting age effect.

\item{} An age effect is confirmed with our mean age indicator,
$\Delta(bz-yz)$,  which finds a correlation with luminosity such that
bright ellipticals are about 2 Gyrs younger than dwarf ellipticals.  This
is a small and subtle difference, only detectable due to the wide range
in luminosity for our combined Coma/Fornax sample.  Since spectroscopic
studies are restricted to the top of the luminosity function, it is
unsurprising that this effect has been missed in previous work.  This
result is in agreement with the results from Terlevich \& Forbes (2001)
who also find that young galaxies (high mass) have high metallicities and
that old galaxies (low mass) have a broad range of metallicities.

\end{itemize}

Interpretation of the above observations is problematic.  A classic
galactic wind model reproduces the CM relation by a simple relationship
between the mass of a galaxy and its ability to retain gas by the depth of
its gravitational well leading to longer, and more chemically evolved,
star formation.  Observations of distant clusters strongly support the
metallicity interpretation for the CM relation (Kodama \& Arimoto 1997).
The wind model also predicts a mean age difference since more massive
galaxies have a longer episode of star formation, whereas dwarf systems
are extinguished quickly producing a relative older mean age.  However,
for traditional models (Bressan, Chiosi \& Fagotto 1994), this age
difference is extremely minor and the age effect for $bz-yz$ remains
unresolved by a wind model.

It seems almost inescapable that the cluster environment plays an
important role to the stellar populations of cluster ellipticals, especially
since the most massive systems have evidence of a history of past
mergers.  While the number of galaxies with blue populations is very low
in present-day clusters, this was not the case in the recent past (i.e.
Butcher-Oemler effect).  Mergers with gas-rich, star forming galaxies
would have a minor effect on the metallicity color, yet would dramatically
lower the mean stellar age as estimated by the $\Delta(bz-yz)$ index.
Thus, later mergers by galaxies such as spirals, with young stellar
populations would explain the younger mean age of massive ellipticals,
which is in agreement with the discovery of a large number of red merger
systems in MS 1054-03 ($z=0.83$, van Dokkum \etal 1999).  This also
provides a cautionary tale for the study of CM relationships in distant
clusters.  If an intermediate age population is visible in present-day
clusters, then at moderate redshifts these same galaxies may not even be
on the CM relation.  For example, the scenarios developed by Kauffmann
(1996) indicate that some ellipticals could enter the red envelope by
redshift of 0.4 to 0.5, even though most of the stars formed at redshift
greater than 2.

Lastly, our results also favor the so-called hierarchical clustering and
merging scenario of galaxy formation (Kauffmann, White \& Guiderdoni 1993, Kauffmann \&
Charlot 1998) where the massive galaxies are assembled from older, lower
mass systems.  Younger mean age and higher metallicities are achieved by
star formation as massive ellipticals consume disk galaxies, and their
cold gas, and efficiently turn this material into the bulk of their
stellar populations.  While the scenario to convert low metallicity,
gas-rich systems into present-day red, massive ellipticals seem finely
tuned to produce the correct metallicities, the difference between
present-day dwarf and giant elliptical ages (about 2 Gyrs, see Figure 5)
does match the predictions of the hierarchical models (see Kauffmann \&
Charlot 1998 Figure 3).  However, the data only supports a conclusion that
massive systems are younger, whether that is from assembly history or an
extended star formation history is unclear.  The narrow range of
metallicity for high mass galaxies argues for a uniform star formation
history in high mass systems.  The range in metallicities for younger
systems implies a more stochastic process where the weaker gravitational
fields are more susceptible to environmental effects such as cluster ram
pressure stripping.

\acknowledgements
The authors wish to thank NOAO and Steward Observatory for granting
telescope time for this project.  One of us (K. Rakos) gratefully
acknowledges the financial support from the Austrian Fonds zur Foerderung
der wissenschaftlichen Forschung.

\clearpage

\begin{deluxetable}{lccccccccccl}
\tablecolumns{12}
\small
\tablewidth{0pt}
\tablecaption{NGC 4889 Group}
\tablehead{
\colhead{object} & 
\multicolumn{2}{c}{R.A. (J2000) Dec.} &
\colhead{$uz-vz$} &
\colhead{$bz-yz$} &
\colhead{$vz-yz$} &
\colhead{$mz$} &
\colhead{$m_{5500}$} &
\colhead{$M_{5500}$} &
\colhead{[Fe/H]} & 
\colhead{$\Delta(bz-yz)$} &
\colhead{Class} \nl
}
\startdata
a001 & 195.11765 & 27.97243 & 0.57 & 0.25 & 0.71 & +0.21 & 14.5 & -20.3 & -0.15 & -0.047 & E \nl
a002 & 195.11595 & 27.95587 & 0.63 & 0.28 & 0.69 & +0.13 & 15.0 & -19.8 & -0.22 & -0.011 & E \nl
a004 & 195.10321 & 27.92628 & 1.14 & 0.26 & 0.55 & +0.03 & 17.1 & -17.7 & -0.75 & +0.008 & E \nl
a005 & 195.10493 & 27.94904 & 1.24 & 0.10 & 0.04 & -0.16 & 19.1 & -15.7 & -2.68 & -0.011 & E \nl
a008 & 195.09177 & 28.04706 & 0.59 & 0.29 & 0.67 & +0.09 & 14.6 & -20.2 & -0.30 & +0.004 & E \nl
a009 & 195.07054 & 28.06402 & 0.34 & 0.24 & 0.63 & +0.15 & 15.5 & -19.3 & ... & ...  & A \nl
a010 & 195.07803 & 28.00925 & 0.79 & 0.32 & 0.54 & -0.10 & 17.4 & -17.4 & -0.79 & +0.070 & E \nl
a011 & 195.07297 & 27.98753 & 0.78 & 0.20 & 0.58 & +0.18 & 17.1 & -17.7 & -0.64 & -0.061 & E \nl
a012 & 195.06839 & 27.96748 & 0.76 & 0.30 & 0.51 & -0.09 & 14.7 & -20.1 & -0.90 & +0.059 & E \nl
a013 & 195.07330 & 27.95535 & 0.71 & 0.33 & 0.70 & +0.04 & 13.7 & -21.1 & -0.18 & +0.036 & E \nl
a014 & 195.07784 & 27.93681 & 0.19 & 0.26 & 0.53 & +0.01 & 15.9 & -18.9 & ... & ...  & A+ \nl
a016 & 195.04771 & 27.90973 & 0.06 & 0.14 & 0.15 & -0.13 & 18.1 & -16.7 & ... & ...  & S+ \nl
a017 & 195.03104 & 27.95817 & 0.60 & 0.39 & 0.63 & -0.15 & 16.9 & -17.9 & ... & ...  & S- \nl
a018 & 195.03491 & 27.95486 & 1.20 & 0.35 & 0.75 & +0.05 & 17.6 & -17.2 & +0.01 & +0.042 & E \nl
a020 & 195.03368 & 27.97692 & 0.87 & 0.37 & 0.75 & +0.01 & 12.1 & -22.7 & +0.01 & +0.062 & E \nl
a021 & 195.06104 & 28.04119 & 0.71 & 0.32 & 0.61 & -0.03 & 14.9 & -19.9 & -0.52 & +0.051 & E \nl
a023 & 195.02196 & 28.02441 & 0.62 & 0.39 & 0.58 & -0.20 & 15.7 & -19.1 & ... & ...  & S- \nl
a025 & 195.02632 & 28.00390 & 0.66 & 0.32 & 0.68 & +0.04 & 15.0 & -19.8 & -0.26 & +0.032 & E \nl
a026 & 195.01826 & 27.98746 & 0.75 & 0.33 & 0.62 & -0.04 & 14.1 & -20.7 & -0.49 & +0.058 & E \nl
a028 & 195.00418 & 27.94552 & 0.64 & 0.20 & 0.31 & -0.09 & 17.7 & -17.1 & ... & ...  & S- \nl
a029 & 194.99791 & 27.94050 & 1.17 & 0.29 & 0.36 & -0.22 & 17.6 & -17.2 & -1.47 & +0.090 & E \nl
a030 & 194.98627 & 27.93001 & 0.68 & 0.34 & 0.69 & +0.01 & 15.5 & -19.3 & -0.22 & +0.049 & E \nl
a031 & 194.95891 & 27.91228 & 0.56 & 0.37 & 0.74 & +0.00 & 16.8 & -18.0 & -0.03 & +0.065 & E \nl
a032 & 194.95872 & 27.92476 & 0.60 & 0.33 & 0.83 & +0.17 & 16.2 & -18.6 & +0.31 & +0.000 & E \nl
a033 & 194.99225 & 27.95422 & 0.98 & 0.25 & 0.32 & -0.18 & 18.1 & -16.7 & -1.62 & +0.061 & E \nl
a034 & 194.97449 & 27.97064 & 0.33 & 0.36 & 0.44 & -0.28 & 17.3 & -17.5 & ... & ...  & S \nl
a035 & 194.94420 & 27.97405 & 0.78 & 0.34 & 0.79 & +0.11 & 14.7 & -20.1 & +0.16 & +0.021 & E \nl
a036 & 194.95183 & 27.98291 & 0.83 & 0.22 & 0.26 & -0.18 & 17.0 & -17.8 & -1.85 & +0.048 & E \nl
a037 & 194.94472 & 27.99217 & 1.15 & 0.29 & 0.52 & -0.06 & 16.8 & -18.0 & -0.86 & +0.046 & E \nl
a038 & 194.94385 & 28.00025 & 0.72 & 0.25 & 0.36 & -0.14 & 17.3 & -17.5 & ... & ...  & S- \nl
a039 & 194.98268 & 28.03465 & 0.76 & 0.33 & 0.67 & +0.01 & 14.4 & -20.4 & -0.30 & +0.044 & E \nl
\enddata
\end{deluxetable}

\begin{deluxetable}{lccccccccccl}
\tablecolumns{12}
\small
\tablewidth{0pt}
\tablecaption{NGC 4874 Group}
\tablehead{
\colhead{object} & 
\multicolumn{2}{c}{R.A. (J2000) Dec.} &
\colhead{$uz-vz$} &
\colhead{$bz-yz$} &
\colhead{$vz-yz$} &
\colhead{$mz$} &
\colhead{$m_{5500}$} &
\colhead{$M_{5500}$} &
\colhead{[Fe/H]} & 
\colhead{$\Delta(bz-yz)$} &
\colhead{Class} \nl
}
\startdata
b041 & 194.93370 & 27.86699 & 1.01 & 0.30 & 0.58 & -0.02 & 15.7 & -19.1 & -0.64 & +0.039 & E \nl
b042 & 194.93550 & 27.88941 & 0.66 & 0.28 & 0.46 & -0.10 & 17.4 & -17.4 & ... & ...  & S- \nl
b043 & 194.95770 & 27.90875 & 0.85 & 0.38 & 0.73 & -0.03 & 17.7 & -17.1 & -0.07 & +0.078 & E \nl
b044 & 194.93430 & 27.91231 & 0.82 & 0.35 & 0.72 & -0.02 & 14.6 & -20.2 & -0.11 & +0.051 & E \nl
b045 & 194.92560 & 27.92468 & 0.79 & 0.34 & 0.67 & -0.01 & 15.6 & -19.2 & -0.30 & +0.054 & E \nl
b046 & 194.93370 & 27.95802 & 0.76 & 0.31 & 0.57 & -0.05 & 15.8 & -19.0 & -0.68 & +0.052 & E \nl
b047 & 194.93179 & 27.99426 & 0.90 & 0.31 & 0.56 & -0.06 & 15.3 & -19.5 & -0.71 & +0.055 & E \nl
b048 & 194.90781 & 28.00071 & 1.12 & 0.27 & 0.47 & -0.07 & 16.9 & -17.9 & -1.05 & +0.040 & E \nl
b049 & 194.90910 & 27.98681 & 0.85 & 0.32 & 0.64 & +0.00 & 15.9 & -18.9 & -0.41 & +0.043 & E \nl
b050 & 194.88609 & 27.98334 & 0.78 & 0.32 & 0.64 & +0.00 & 14.6 & -20.2 & -0.41 & +0.043 & E \nl
b051 & 194.91730 & 27.96798 & 0.74 & 0.34 & 0.71 & +0.03 & 16.1 & -18.7 & -0.15 & +0.043 & E \nl
b052 & 194.91460 & 27.95369 & 0.67 & 0.34 & 0.73 & +0.05 & 15.8 & -19.0 & -0.07 & +0.038 & E \nl
b053 & 194.89160 & 27.94679 & 0.76 & 0.38 & 0.77 & +0.01 & 14.7 & -20.1 & +0.08 & +0.067 & E \nl
b054 & 194.89600 & 27.93476 & 0.95 & 0.26 & 0.29 & -0.23 & 17.9 & -16.9 & -1.73 & +0.080 & E \nl
b055 & 194.89720 & 27.90611 & 0.90 & 0.41 & 0.61 & -0.21 & 18.3 & -16.5 & -0.52 & +0.141 & E \nl
b056 & 194.90710 & 27.90736 & 0.72 & 0.36 & 0.77 & +0.05 & 14.9 & -19.9 & +0.08 & +0.047 & E \nl
b057 & 194.87750 & 27.88416 & 0.92 & 0.35 & 0.64 & -0.06 & 14.3 & -20.5 & -0.41 & +0.073 & E \nl
b058 & 194.88310 & 27.92120 & 0.89 & 0.35 & 0.47 & -0.23 & 18.0 & -16.8 & -1.05 & +0.120 & E \nl
b059 & 194.87151 & 27.94231 & 0.86 & 0.32 & 0.65 & +0.01 & 18.7 & -16.1 & -0.37 & +0.040 & E \nl
b060 & 194.87910 & 27.95481 & 1.01 & 0.37 & 0.48 & -0.26 & 17.7 & -17.1 & -1.02 & +0.137 & E \nl
b061 & 194.87421 & 27.95642 & 0.83 & 0.36 & 0.75 & +0.03 & 14.6 & -20.2 & +0.01 & +0.052 & E \nl
b062 & 194.88370 & 28.00196 & 0.59 & 0.35 & 0.47 & -0.23 & 18.6 & -16.2 & ... & ...  & S- \nl
b063 & 194.90390 & 28.01830 & 0.37 & 0.28 & 0.65 & +0.09 & 17.8 & -17.0 & ... & ...  & A \nl
b064 & 194.88049 & 28.04687 & 0.83 & 0.32 & 0.70 & +0.06 & 15.1 & -19.7 & -0.18 & +0.026 & E \nl
b065 & 194.86890 & 28.04081 & 0.89 & 0.31 & 0.64 & +0.02 & 15.7 & -19.1 & -0.41 & +0.033 & E \nl
b066 & 194.86819 & 28.01925 & 0.55 & 0.25 & 0.53 & +0.03 & 18.4 & -16.4 & ... & ...  & S- \nl
b067 & 194.87300 & 28.00630 & 0.53 & 0.24 & 0.19 & -0.29 & 19.5 & -15.3 & ... & ...  & S \nl
b068 & 194.86031 & 27.99831 & 0.70 & 0.34 & 0.64 & -0.04 & 16.0 & -18.8 & -0.41 & +0.063 & E \nl
b069 & 194.85381 & 27.99657 & 0.74 & 0.34 & 0.61 & -0.07 & 17.1 & -17.7 & -0.52 & +0.071 & E \nl
b070 & 194.85561 & 27.97312 & 0.82 & 0.33 & 0.68 & +0.02 & 15.5 & -19.3 & -0.26 & +0.042 & E \nl
b071 & 194.85500 & 27.96792 & 0.82 & 0.34 & 0.71 & +0.03 & 15.8 & -19.0 & -0.15 & +0.043 & E \nl
b072 & 194.85500 & 27.93464 & 0.83 & 0.32 & 0.55 & -0.09 & 17.0 & -17.8 & -0.75 & +0.068 & E \nl
b073 & 194.84689 & 27.91965 & 0.78 & 0.36 & 0.76 & +0.04 & 17.2 & -17.6 & +0.04 & +0.049 & E \nl
b075 & 194.84650 & 27.91174 & 0.75 & 0.35 & 0.76 & +0.06 & 13.9 & -20.9 & +0.04 & +0.039 & E \nl
b076 & 194.84860 & 27.89026 & 0.74 & 0.23 & 0.21 & -0.25 & 18.3 & -16.5 & ... & ...  & S- \nl
b077 & 194.84340 & 27.89692 & 0.77 & 0.32 & 0.58 & -0.06 & 15.5 & -19.3 & -0.64 & +0.059 & E \nl
b078 & 194.83299 & 27.88579 & 0.77 & 0.29 & 0.56 & -0.02 & 15.5 & -19.3 & -0.71 & +0.035 & E \nl
b079 & 194.80960 & 27.89528 & 0.87 & 0.43 & 0.64 & -0.22 & 17.5 & -17.3 & -0.41 & +0.153 & E \nl
b080 & 194.81371 & 27.89891 & 1.13 & 0.37 & 0.64 & -0.10 & 18.3 & -16.5 & -0.41 & +0.093 & E \nl
b081 & 194.79680 & 27.98085 & 0.82 & 0.19 & 0.15 & -0.23 & 17.0 & -17.8 & ... & ...  & S- \nl
b082 & 194.83121 & 27.96808 & 0.92 & 0.34 & 0.68 & +0.00 & 17.8 & -17.0 & -0.26 & +0.052 & E \nl
b083 & 194.83859 & 27.97356 & 0.88 & 0.29 & 0.52 & -0.06 & 16.5 & -18.3 & -0.86 & +0.046 & E \nl
b084 & 194.83141 & 27.97346 & 1.18 & 0.29 & 0.54 & -0.04 & 17.5 & -17.3 & -0.79 & +0.040 & E \nl
b085 & 194.81261 & 27.97074 & 0.91 & 0.33 & 0.63 & -0.03 & 14.6 & -20.2 & -0.45 & +0.055 & E \nl
b086 & 194.80341 & 27.97701 & 0.86 & 0.33 & 0.79 & +0.13 & 13.9 & -20.9 & +0.16 & +0.011 & E \nl
b087 & 194.81870 & 27.98257 & 0.45 & 0.34 & 0.64 & -0.04 & 19.8 & -15.0 & ... & ...  & A- \nl
b088 & 194.79649 & 28.00965 & 0.92 & 0.39 & 0.67 & -0.11 & 17.2 & -17.6 & -0.30 & +0.104 & E \nl
b089 & 194.89841 & 27.95914 & 0.88 & 0.36 & 0.78 & +0.06 & 12.1 & -22.7 & +0.12 & +0.044 & E \nl
\enddata
\end{deluxetable}

\begin{deluxetable}{lccccccccccl}
\tablecolumns{12}
\small
\tablewidth{0pt}
\tablecaption{NGC 4839 Group}
\tablehead{
\colhead{object} & 
\multicolumn{2}{c}{R.A. (J2000) Dec.} &
\colhead{$uz-vz$} &
\colhead{$bz-yz$} &
\colhead{$vz-yz$} &
\colhead{$mz$} &
\colhead{$m_{5500}$} &
\colhead{$M_{5500}$} &
\colhead{[Fe/H]} & 
\colhead{$\Delta(bz-yz)$} &
\colhead{Class} \nl
}
\startdata
c090 & 194.29520 & 27.56129 & 0.61 & 0.37 & 0.61 & -0.13 & 18.4 & -16.4 & ... & ...  & S- \nl
c091 & 194.30000 & 27.58170 & 0.59 & 0.23 & 0.06 & -0.40 & 19.5 & -15.3 & ... & ...  & S \nl
c092 & 194.34760 & 27.55026 & 0.82 & 0.33 & 0.59 & -0.07 & 17.3 & -17.5 & -0.60 & +0.067 & E \nl
c093 & 194.35710 & 27.54647 & 0.90 & 0.38 & 0.70 & -0.06 & 14.8 & -20.0 & -0.18 & +0.086 & E \nl
c094 & 194.37691 & 27.54337 & 0.77 & 0.37 & 0.80 & +0.06 & 17.1 & -17.7 & +0.19 & +0.048 & E \nl
c095 & 194.38860 & 27.54876 & 0.64 & 0.14 & -0.02 & -0.30 & 17.5 & -17.3 & ... & ...  & S \nl
c096 & 194.39600 & 27.56802 & 0.67 & 0.41 & 0.63 & -0.21 & 19.7 & -15.1 & -0.45 & +0.135 & E \nl
c097 & 194.42931 & 27.57742 & 0.75 & 0.34 & 0.70 & +0.02 & 16.5 & -18.3 & -0.18 & +0.046 & E \nl
c098 & 194.44240 & 27.57191 & 0.94 & 0.38 & 0.76 & +0.00 & 20.0 & -14.8 & +0.04 & +0.069 & E \nl
c099 & 194.46870 & 27.57764 & 0.83 & 0.35 & 0.75 & +0.05 & 19.5 & -15.3 & +0.01 & +0.042 & E \nl
c100 & 194.46181 & 27.53464 & 0.77 & 0.26 & 0.30 & -0.22 & 18.8 & -16.0 & ... & ...  & S- \nl
c101 & 194.45911 & 27.52099 & 0.54 & 0.22 & 0.29 & -0.15 & 19.6 & -15.2 & ... & ...  & S \nl
c102 & 194.35130 & 27.49843 & 0.90 & 0.39 & 0.84 & +0.06 & 12.4 & -22.4 & +0.35 & +0.057 & E \nl
c103 & 194.36481 & 27.53753 & 0.69 & 0.34 & 0.62 & -0.06 & 18.5 & -16.3 & -0.49 & +0.068 & E \nl
c104 & 194.31120 & 27.51770 & 0.22 & 0.28 & 0.45 & -0.11 & 21.2 & -13.6 & ... & ...  & A+ \nl
c105 & 194.28650 & 27.50831 & 0.59 & 0.38 & 0.47 & -0.29 & 20.6 & -14.2 & ... & ...  & S- \nl
c106 & 194.29530 & 27.52918 & 0.76 & 0.35 & 0.68 & -0.02 & 19.2 & -15.6 & -0.26 & +0.062 & E \nl
c107 & 194.26750 & 27.52678 & 0.78 & 0.34 & 0.68 & +0.00 & 16.5 & -18.3 & -0.26 & +0.052 & E \nl
c108 & 194.30521 & 27.54821 & 0.54 & 0.40 & 0.58 & -0.22 & 20.1 & -14.7 & ... & ...  & S- \nl
c109 & 194.23320 & 27.48264 & 0.59 & 0.38 & 0.61 & -0.15 & 20.5 & -14.3 & ... & ...  & S- \nl
c110 & 194.23340 & 27.48046 & 0.33 & 0.33 & 0.67 & +0.01 & 18.2 & -16.6 & ... & ...  & A \nl
c111 & 194.23241 & 27.46325 & 0.69 & 0.35 & 0.70 & +0.00 & 17.5 & -17.3 & -0.18 & +0.056 & E \nl
c112 & 194.26241 & 27.46611 & 0.50 & 0.22 & 0.30 & -0.14 & 19.9 & -14.9 & ... & ...  & S \nl
c113 & 194.26199 & 27.46325 & 0.65 & 0.32 & 0.58 & -0.06 & 20.0 & -14.8 & -0.64 & +0.059 & E \nl
c114 & 194.28900 & 27.46725 & 0.92 & 0.39 & 0.85 & +0.09 & 14.3 & -20.5 & +0.38 & +0.055 & E \nl
c115 & 194.28430 & 27.49088 & 0.85 & 0.34 & 0.55 & -0.13 & 20.6 & -14.2 & -0.75 & +0.088 & E \nl
c116 & 194.34039 & 27.47550 & 0.82 & 0.36 & 0.71 & -0.01 & 17.1 & -17.7 & -0.15 & +0.063 & E \nl
c117 & 194.34200 & 27.47079 & 0.52 & 0.41 & 0.67 & -0.15 & 20.2 & -14.6 & ... & ...  & A- \nl
c118 & 194.35060 & 27.46414 & 0.47 & 0.35 & 0.48 & -0.22 & 20.7 & -14.1 & ... & ...  & S \nl
c119 & 194.40070 & 27.48488 & 0.88 & 0.40 & 0.86 & +0.06 & 15.3 & -19.5 & +0.42 & +0.062 & E \nl
c120 & 194.39940 & 27.49360 & 0.92 & 0.41 & 0.90 & +0.08 & 14.2 & -20.6 & +0.57 & +0.061 & E \nl
c121 & 194.44971 & 27.49450 & 0.42 & 0.24 & 0.52 & +0.04 & 19.0 & -15.8 & ... & ...  & S \nl
c122 & 194.46181 & 27.49094 & 0.96 & 0.38 & 0.73 & -0.03 & 17.7 & -17.1 & -0.07 & +0.078 & E \nl
c123 & 194.47690 & 27.49059 & 0.89 & 0.29 & 0.51 & -0.07 & 15.3 & -19.5 & -0.90 & +0.049 & E \nl
c124 & 194.46851 & 27.47190 & 0.53 & 0.10 & -0.10 & -0.30 & 19.0 & -15.8 & ... & ...  & S+ \nl
c125 & 194.45790 & 27.47672 & 0.51 & 0.34 & 0.36 & -0.32 & 19.1 & -15.7 & ... & ...  & S \nl
c126 & 194.47279 & 27.45355 & 1.07 & 0.26 & 0.58 & +0.06 & 20.0 & -14.8 & -0.64 & -0.001 & E \nl
c127 & 194.46770 & 27.44369 & 0.84 & 0.24 & 0.44 & -0.04 & 19.8 & -15.0 & -1.17 & +0.018 & E \nl
c128 & 194.48730 & 27.44220 & 0.73 & 0.12 & 0.00 & -0.24 & 17.5 & -17.3 & ... & ...  & S \nl
c129 & 194.44150 & 27.42925 & 0.75 & 0.28 & 0.58 & +0.02 & 16.4 & -18.4 & -0.64 & +0.019 & E \nl
c130 & 194.45219 & 27.45402 & 0.97 & 0.19 & 0.24 & -0.14 & 18.4 & -16.4 & -1.92 & +0.024 & E \nl
c131 & 194.43280 & 27.46159 & 0.64 & 0.27 & 0.55 & +0.01 & 19.3 & -15.5 & ... & ...  & S- \nl
c132 & 194.40370 & 27.46332 & 0.40 & 0.08 & 0.13 & -0.03 & 20.1 & -14.7 & ... & ...  & S+ \nl
c133 & 194.41180 & 27.45552 & 0.83 & 0.33 & 0.56 & -0.10 & 18.2 & -16.6 & -0.71 & +0.075 & E \nl
c134 & 194.30270 & 27.45796 & 0.82 & 0.27 & 0.37 & -0.17 & 19.9 & -14.9 & -1.43 & +0.068 & E \nl
c135 & 194.24809 & 27.45603 & 0.76 & 0.30 & 0.47 & -0.13 & 19.2 & -15.6 & -1.05 & +0.070 & E \nl
c136 & 194.25661 & 27.37173 & 0.88 & 0.36 & 0.82 & +0.10 & 15.5 & -19.3 & +0.27 & +0.033 & E \nl
c137 & 194.26350 & 27.38813 & 0.75 & 0.35 & 0.51 & -0.19 & 20.0 & -14.8 & -0.90 & +0.109 & E \nl
c138 & 194.30119 & 27.38709 & 0.93 & 0.25 & 0.40 & -0.10 & 19.2 & -15.6 & -1.32 & +0.039 & E \nl
c139 & 194.29480 & 27.40521 & 0.81 & 0.34 & 0.67 & -0.01 & 15.3 & -19.5 & -0.30 & +0.054 & E \nl
c140 & 194.29660 & 27.40131 & 0.64 & 0.37 & 0.55 & -0.19 & 19.7 & -15.1 & ... & ...  & S- \nl
c141 & 194.29410 & 27.41943 & 0.53 & 0.27 & 0.33 & -0.21 & 20.7 & -14.1 & ... & ...  & S \nl
c142 & 194.30521 & 27.41071 & 0.72 & 0.29 & 0.59 & +0.01 & 18.8 & -16.0 & -0.60 & +0.027 & E \nl
c143 & 194.30679 & 27.40899 & 0.45 & 0.30 & 0.48 & -0.12 & 20.0 & -14.8 & ... & ...  & S \nl
c144 & 194.33990 & 27.39591 & 0.53 & 0.36 & 0.61 & -0.11 & 20.6 & -14.2 & ... & ...  & S- \nl
c145 & 194.35190 & 27.38834 & 0.40 & 0.43 & 0.69 & -0.17 & 19.0 & -15.8 & ... & ...  & A \nl
c146 & 194.35580 & 27.40462 & 0.66 & 0.24 & 0.22 & -0.26 & 15.5 & -19.3 & ... & ...  & S \nl
c147 & 194.35880 & 27.40519 & 0.74 & 0.36 & 0.59 & -0.13 & 16.8 & -18.0 & -0.60 & +0.097 & E \nl
c148 & 194.40610 & 27.41045 & 0.94 & 0.22 & 0.44 & +0.00 & 19.4 & -15.4 & -1.17 & -0.002 & E \nl
c149 & 194.48770 & 27.38107 & 0.89 & 0.33 & 0.66 & +0.00 & 17.0 & -17.8 & -0.34 & +0.047 & E \nl
\enddata
\end{deluxetable}

\begin{deluxetable}{lrrrrrrrr}
\tablecolumns{9}
\small
\tablewidth{0pt}
\tablecaption{Cluster Population Fractions}
\tablehead{
\colhead{} &
\multicolumn{2}{c}{NGC 4889} &
\multicolumn{2}{c}{NGC 4874} &
\multicolumn{2}{c}{NGC 4839} &
\multicolumn{2}{c}{Total} \nl
}
\startdata
E    & 23 & (74\%) & 40 & (83\%) & 35 & (58\%) & 98 & (71\%) \nl
S-   &  4 & (13\%) &  5 & (10\%) &  8 & (13\%) & 17 & (12\%) \nl
S    &  1 & ( 3\%) &  1 & ( 2\%) & 11 & (18\%) & 13 & ( 9\%) \nl
S+   &  1 & ( 3\%) &  0 & ( 0\%) &  2 & ( 3\%) &  3 & ( 2\%) \nl
A    &  2 & ( 6\%) &  2 & ( 4\%) &  4 & ( 7\%) &  8 & ( 6\%) \nl
     & 31 & (22\%) & 48 & (35\%) & 60 & (43\%) &139 & \nl
     & & & & & & & & \nl
blue &  5 & (16\%) &  1 & ( 2\%) &  8 & (13\%) & 14 & (10\%) \nl
red  & 26 & (84\%) & 47 & (98\%) & 52 & (87\%) &125 & (90\%) \nl
\enddata
\end{deluxetable}

\clearpage

\end{document}